\documentclass[11pt,twoside]{article}
\usepackage{./asp2014}
\usepackage{fixltx2e}

\aspSuppressVolSlug
\resetcounters

\begin{document}

\title{Discovery of the Earliest-Type Oe Stars in the \\Small Magellanic Cloud}
\author{Jesse B. Golden-Marx$^1$, M.S. Oey$^2$, Joel B. Lamb$^3$, and Andrew S. Graus$^4$}
\affil{$^1$Department of Astronomy, University of Michigan, Ann Arbor, MI, 48109, USA; \email{jessegm@umich.edu}}
\affil{$^2$Department of Astronomy, University of Michigan, Ann Arbor, MI, 48109, USA}
\affil{$^3$Department of Physical Sciences, Nassau Community College, Garden City, NY, 11530, USA }
\affil{$^4$Department of Physics and Astronomy, University of California, Irvine, Irvine, CA, 92697, USA }

\paperauthor{Jesse B. Golden-Marx}{jessegm@umich.edu}{ORCID_Or_Blank}{University of Michigan}{Department of Astronomy}{Ann Arbor}{MI}{48104}{USA}
\paperauthor{M.S. Oey}{msoey@umich.edu}{ORCID_Or_Blank}{University of Michigan}{Department of Astronomy}{Ann Arbor}{MI}{48104}{USA}
\paperauthor{Joel B. Lamb}{Joel.Lamb@ncc.edu}{ORCID_Or_Blank}{Nassau Community College}{Department of Physical Sciences}{Garden City}{New York}{11530}{USA}
\paperauthor{Andrew S. Graus}{agraus@uci.edu}{ORCID_Or_Blank}{University of California, Irvine}{Department of Physics and Astronomy}{Irvine}{California}{92697}{USA}

\begin{abstract}
No classical Oe/Be stars with spectral type earlier than O7.5e have been identified to date in the Milky Way (MW).  This is consistent with the decretion disk model because strong stellar winds cause early-type O stars to lose angular momentum, thereby preventing them from rotating fast enough to spin out decretion disks.  However, metal-poor O stars have weaker stellar winds, allowing the stars to retain angular momentum.  Therefore, low-metallicity environments should promote the formation of Oe stars, including those of earlier spectral types than observed in high-metallicity environments.  Using the RIOTS4 survey, a spatially complete spectroscopic survey of Small Magellanic Cloud (SMC) field OB stars taken with the IMACS multi-slit spectrograph at the Magellan Baade Telescope, we identify 25-31 SMC field Oe stars, which account for 20-28\% of SMC field O stars.  This fraction is significantly higher than in the MW, where < 10-15\% of O stars display the Be phenomenon.  We also present 5-7 Oe stars of spectral type ranging from O5.5e to O7e, all earlier spectral types than the earliest MW Oe star.  These early type Oe stars represent 20-23\% of our SMC Oe stars, a dramatic increase compared to the MW, where no Oe stars have been identified with these early spectral types.  Thus, the higher frequencies of Oe stars and their earlier spectral range in the metal-poor SMC are consistent with the decretion disk model.    
\end{abstract}

\section{Introduction: Oe Stars} 
  
In previous sessions, we discussed the characteristics of Be stars and their decretion disks.  According to the standard model, a decretion disk forms when the rapidly rotating star ejects its outer layers to shed angular momentum as the star rotates near the critical velocity (e.g., Struve 1931), the velocity where the gravitational force equals the centrifugal force.  Our work focuses on observations of Oe stars, a less well understood, early-type extension of the Be phenomenon (e.g., Conti \& Leep 1974), which offer a unique parameter space for us to test the decretion disk model.

While Be stars are common in the Milky Way (MW), Oe stars are rare, particularly earlier-type Oe stars.  Negueruela et al. (2004) identify only 6 Oe stars in a recent survey that includes all previously identified emission line O stars and peculiar O stars visible at La Palma Observatory.  This sample includes the earliest MW Oe star identified, an O7.5 IIIe star, with the majority having later spectral types, between O9 and B0.  While the Negueruela et al. (2004) sample is not statistically complete, it does highlight the relative scarcity of earlier-type Oe stars in the MW when compared to recent surveys identifying hundreds of MW Be stars.  In the MW, the Oe/O ratio, where the number of O stars includes all regular O stars and Oe stars, is estimated to be $\leq$0.10-0.15 for O stars of all spectral types (Negueruela et al. 2004) and observed to be 0.14 for MW cluster O8 -- O9 stars (Martayan et al. 2010).     

Based on the decretion disk model for Oe/Be star formation, the star's ability to retain angular momentum and rotate rapidly is key to forming a decretion disk.  Stellar winds are one of the primary mechanisms for removing angular momentum from a star; however, wind strength is metallicity-dependent (e.g., Kudritzki \& Puls 2000).  In the high-metallicity MW, OB stars have strong winds, which strip angular momentum from the stars, thereby slowing the stellar rotation.  Since O stars have stronger winds than B stars, this can explain the scarcity of MW Oe stars relative to Be stars.  In contrast, low-metallicity O stars have weaker stellar winds, which allow O stars to retain angular momentum and rotate closer to their critical velocity.  Therefore, on average, OB stars should rotate faster in low-metallicity environments than in high-metallicity environments and more O stars should rotate fast enough to form decretion disks, increasing the frequency of Oe/Be stars.  We would also expect earlier type O stars to form decretion disks than those in high metallicity environments.

As discussed by Martayan (these Proceedings), recent observations of OB and Oe/Be stars show how the SMC's low metallicity affects the rotation rate of OB stars and the frequency of Oe/Be stars.  In the MW, OB stars rotate on average at 0.30 -- 0.40~$\Omega_{\rm crit}$, the critical rotational velocity, while in the SMC, the average rotation rate jumps to 0.58~$\Omega_{\rm crit}$ (Martayan et al. 2007).  Thus, on average, SMC OB stars rotate faster than their MW counterparts.  Since the average rotation rate of OB stars increases in low-metallicity environments, the decretion disk model predicts that the frequency of Oe and Be stars should be higher in low-metallicity environments.  Martayan et al. (2010) indeed found that the frequency of Oe/Be stars increases by a factor of 3-5 between the SMC and the MW.  Oe/Be stars are also observed to rotate more rapidly in low-metallicity environments.  In the MW, Be stars rotate on average at 0.84~$\Omega_{\rm crit}$, with a range of 0.69 -- 0.94~$\Omega_{\rm crit}$ (Porter 1996; Cranmer 2005), whereas in the SMC, Be stars rotate at 0.94 -- 1.0~$\Omega_{\rm crit}$ (Martayan et al. 2007).  Based on the observed rotation rate of Be stars in the SMC, LMC, and MW, Martayan et al. (2007) estimate that the minimum rotation rate needed to form decretion disks is approximately 0.70~$\Omega_{\rm crit}$.     

The field or cluster environment may also impact OB star rotation rates.  Observations show that field OB stars rotate more slowly than cluster stars (e.g., Guthrie et al. 1984; Wolff et al. 2007; Wolff et al. 2008).  The slower rotation may result from star formation conditions: field stars may form in lower density environments with lower turbulent velocities and lower infall rates, which would lead to slower angular rotation when the stars form (Wolffet al. 2007; Wolff et al. 2008).  Alternatively, clusters lose stars over time, and field stars would on average be older stars that have dispersed from their clusters.  It is expected that as stars age, their rotation rate decreases because their moment of inertia and radius increase (Huang \& Gies 2010; Huang et al. 2008).  Based on the observations of slowly rotating field stars and the decretion disk model, Oe/Be stars should be more frequent in clusters than in the field.   

\section{Data: The RIOTS4 Survey}

The Runaways and Isolated O Type Star Spectroscopic Survey of the SMC (RIOTS4) is a spatially complete, spectroscopic survey of 374 SMC field OB stars (Lamb et al. 2013).  This survey allows us to test the decretion disk model by determining how metallicity and field versus cluster environment alter the frequency of Oe/Be stars.  The RIOTS4 survey used the photometric criteria of Oey et al. (2004) to identify candidates having \textit{M} > 10$\rm M_{\odot}$: $B\leq15.21$ and $Q_{UBR}\leq-0.84$.  $Q_{UBR}$ is a reddening free parameter, used to select the bluest stars given by,
	\begin{equation}
	Q_{UBR}=(m_{U}-m_{R})-1.396(m_{B}-m_{R}) \quad .
	\end{equation}
In Equation 1, $m_{U}, m_{R},$ and $m_{B}$ are the apparent magnitudes in the \textit{U}, \textit{R}, and \textit{B} bands respectively.  $Q_{UBR}$ is sensitive to the \textit{R} band, and the \textit{R} band is sensitive to H$\alpha$.  Since Oe/Be stars are characterized by strong H$\alpha$ emission, $Q_{UBR}$ enhanced the selection of Oe/Be stars in the RIOTS4 survey.  To identify the field stars, the RIOTS4 survey used a friends-of-friends algorithm that identified field stars as those being more than one clustering length, in this case 28pc, from any other OB star (Lamb et al. 2013; Oey et al. 2004).   

Observations for the RIOTS4 survey were primarily taken with the Inamori Magellan Areal Camera and Spectrograph (IMACS), a multi-slit spectrograph on the Magellan Baade Telescope (Dressler et al. 2006) between 2006 and 2011.  Each observation has a resolution of either R=2700 or 3600.  The majority of the spectra include the wavelength range 4000-5000 \AA~(Lamb et al. 2013), which allows us to use the H$\beta$ emission line to identify Oe/Be stars.     

\section{Spectral Classification}

To classify our stars, we used the classification systems of Walborn \& Fitzpatrick (1990) and Walborn (2009).  To determine the spectral type of our stars we used the ratio between the He~\textsc{i}~$\lambda$4471 and He~\textsc{ii}~$\lambda$4542 absorption line strengths.  We also used the absence of He~\textsc{i}~$\lambda$4144 and He~\textsc{i}~$\lambda$4387 to identify Oe stars with a spectral type earlier than the spectral range O6 -- 6.5.  The equivalent width of the He~\textsc{ii}~$\lambda$4200 absorption line decreases towards later type O stars, so we used the weakest He~\textsc{ii}~$\lambda$4200 absorption lines to identify late-type O stars in the range of O9 -- 9.5 (Walborn \& Fitzpatrick 1990).  Each spectrum was classified independently by three of us (JBGM, MSO, JBL). We then discussed our different classifications to determine the best spectral type.    

One of the primary difficulties in classifying our SMC OB stars is that we cannot use the metal absorption lines to identify later type O and earlier type B stars.  Metal absorption lines, such as Si~\textsc{iv}~$\lambda\lambda$4089,4116, Si~\textsc{iii}~$\lambda$4552, or C~\textsc{iii}~$\lambda$4650, that are present in the spectra of high-metallicity OB stars, are absent in our spectra of low-metallicity SMC OB stars.  Another, equally important, difficulty in classifying our Oe/Be stars arises from infilling of the He~\textsc{i}~$\lambda$4471 absorption line by Be star disk emission (Steele et al. 1999).  Infill in the He~\textsc{i}~$\lambda$4471 absorption line decreases the equivalent width of He~\textsc{i}~$\lambda$4471 and increases the He~\textsc{ii}~$\lambda$4542 /He~\textsc{i}~$\lambda$4471 ratio used to determine the spectral classification of O stars, shifting a star's spectral classification to an earlier type.  Figure 1 shows two distributions of the frequency of Oe/Be spectral classifications in the Negueruela et al. (2004) sample; the purple distribution attempts to correct for infill, while the green distribution does not.   
\articlefigure[scale=.5]{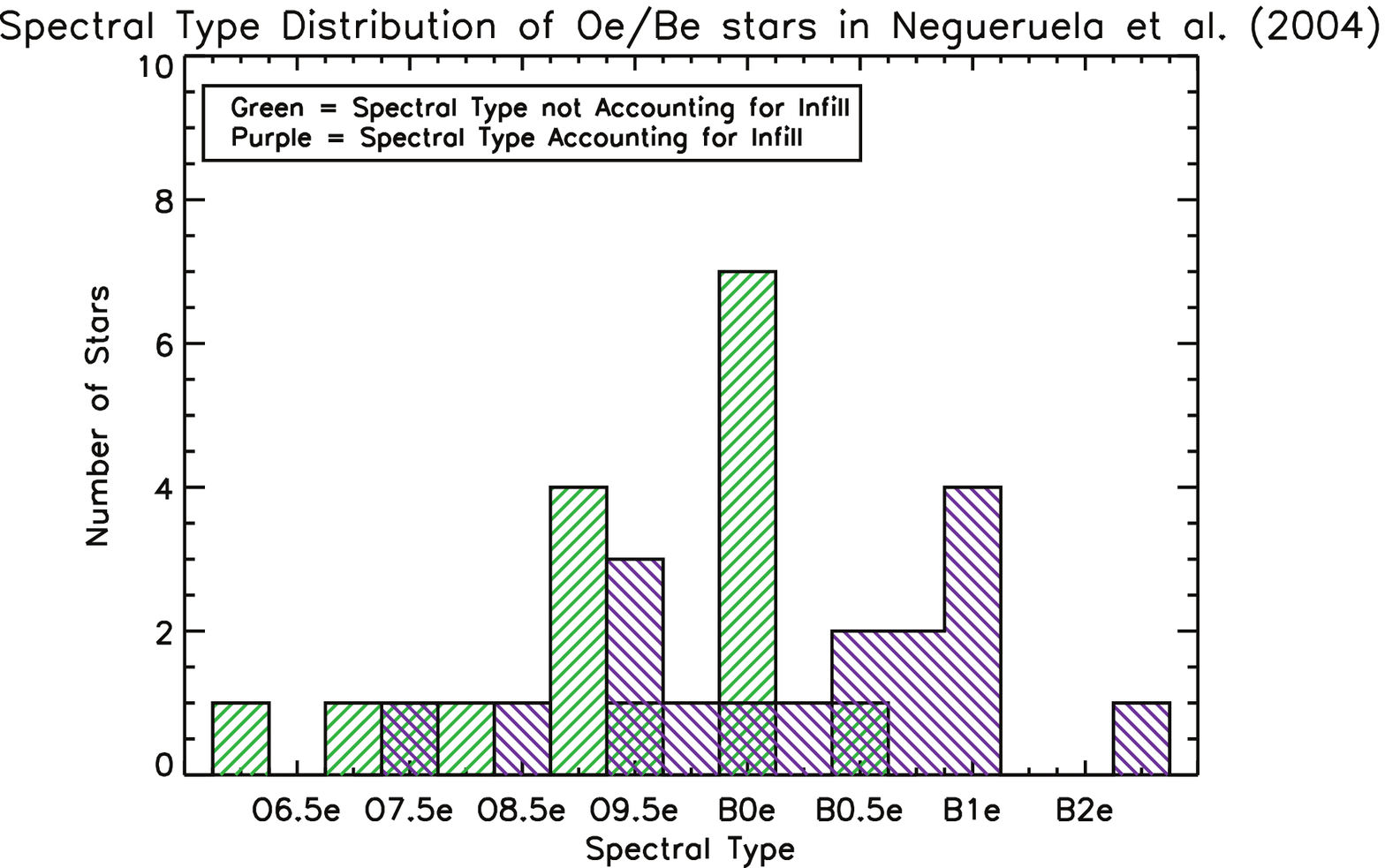}{ex_Figure 1}{The distributions of the spectral types of Oe/Be stars from Negueruela et al. (2004).  The green and purple distributions plot the spectral type of Oe/Be stars when not correcting for infill, and correcting for infill, respectively.}   
Accounting for infill caused the classifications given by Negueruela et al. (2004) to be of later spectral types than previous classifications of the same stars.  To account for infill in our own classifications, we used the presence of He~\textsc{ii}~$\lambda$4200, He~\textsc{i}~$\lambda$4144, and He~\textsc{i}~$\lambda$4387 absorption lines to aid in our classification instead of just relying on the He~\textsc{i}~$\lambda$4471/He~\textsc{ii}~$\lambda$4542 ratio.  However, we caution that this effect causes some uncertainty in our classifications.

\section{Results: The Earliest Oe Stars}

From the RIOTS4 survey, we identified 5 Oe stars with spectral types earlier than O7.5e, the earliest Oe star identified to date in the MW (Negueruela et al. 2004).  The spectra of these 5 stars are shown in Figure 2.  To identify our stars, we use the stellar identification numbers introduced by Massey (2002).  In Figure 2, each star has stronger He~\textsc{ii}~$\lambda$4542 than He~\textsc{i}~$\lambda$4471 absorption, representative of stars of spectral type earlier than O7.5.    
\articlefigure[scale=.45]{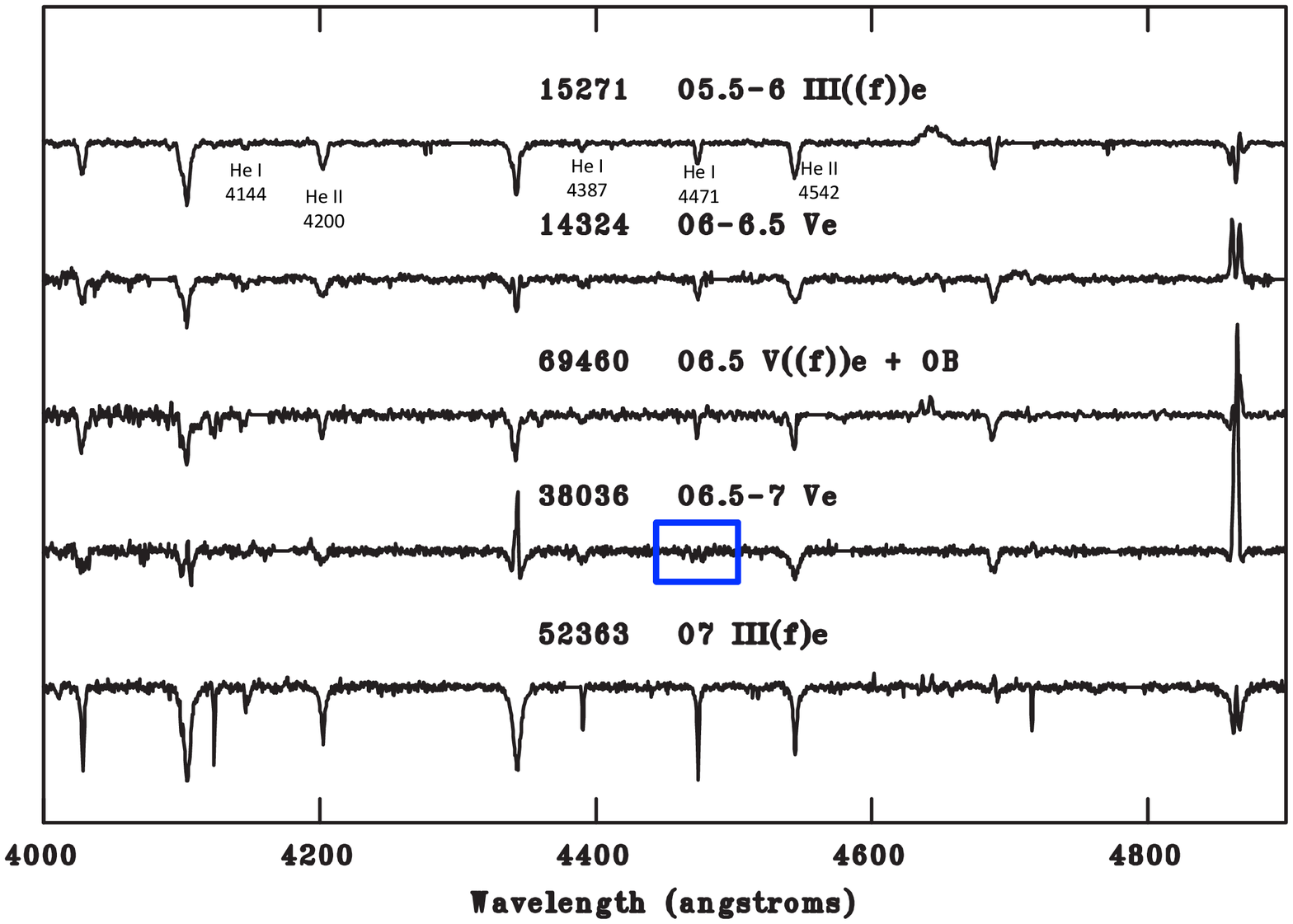}{ex_Figure 2}{The five Oe stars with spectral classifications earlier than O7.5e.  The Massey ID numbers (Massey 2002) and our spectral types are listed above each spectrum.  The major spectral features in the range of 4000 -- 4900\AA \thinspace used for our classifications are labeled.  The He~\textsc{ii}~$\lambda$4542 to He~\textsc{i}~$\lambda$4471 ratio decreases towards later spectral type.  In star 38036, the blue rectangle shows an infilled He~\textsc{i}~$\lambda$4471 absorption line.} 
Our earliest Oe star, number 15271, is an O5.5-6 III((f))e star.  This star displays the He~\textsc{ii}~$\lambda$4542/He~\textsc{i}~$\lambda$4471 absorption line ratio characteristic of an O5.5 star and also the weak He~\textsc{i}~$\lambda$4144 and He~\textsc{i}~$\lambda$4387 absorption of an O6 star.  Star 38036 in Figure 2 displays He~\textsc{i}~$\lambda$4471 infill by the Be star disk emission.  We can see that the He~\textsc{i}~$\lambda$4471 absorption line of this star has a small equivalent width, and is much weaker than He~\textsc{ii}~$\lambda$4542.  While star 38036 displays He~\textsc{i}~$\lambda$4471 infill, we estimate its spectral type based on the presence of weak He~\textsc{i}~$\lambda$4144 and He~\textsc{i}~$\lambda$4387 absorption lines.  While the 5 Oe stars shown in Figure 2 are the earliest classifiable Oe stars in the RIOTS4 survey, we also observed some additional Oe stars that appear to be even hotter and more massive than our O5.5-6e star.  These stars contain He~\textsc{i} emission lines in their spectra, which prevented us from determining accurate spectral types.         

We classified 25 -- 31 Oe stars in the RIOTS4 survey.  The number of Oe stars is given by this range because for a few stars, the classifications are uncertain, and we were only able to determine their spectral types to within a range of values which span the boundary between Oe and Be stars.  Figure 3 shows two distributions of our Oe spectral types from the RIOTS4 survey, one compiled with the earliest spectral classifications for stars with uncertain spectral types, and another with the latest spectral classifications. 
\articlefigure[scale=.45]{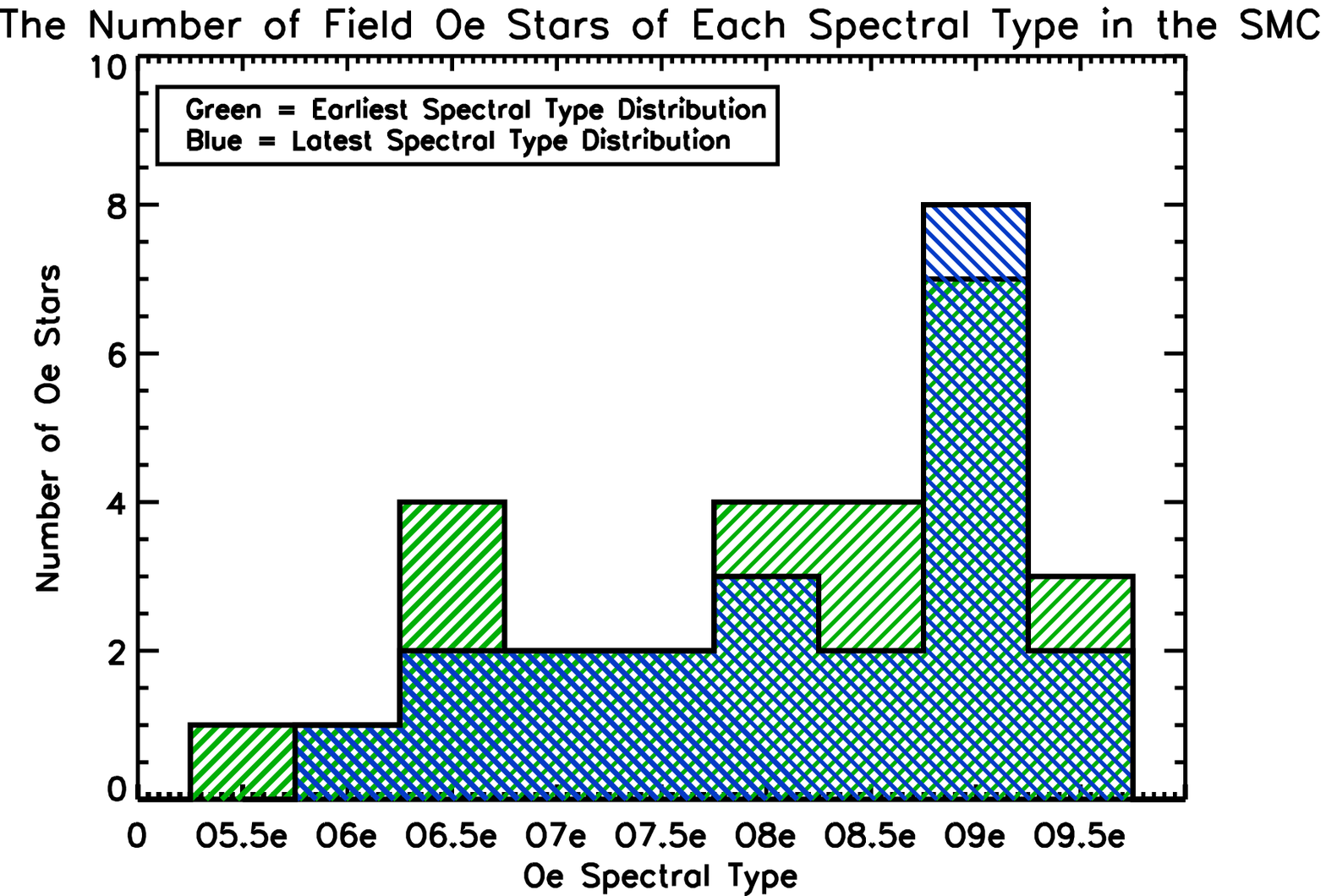}{ex_Figure 3}{Spectral type distributions for our Oe stars.  The green and blue distributions are compiled with the earliest and latest spectral type limits, respectively.}   
Among the RIOTS4 Oe stars are 5 -- 7 stars that are of earlier spectral type than O7.5e, as shown in Figure 3.  Thus, approximately 20 -- 23\% of SMC field Oe stars have earlier spectral types than what is currently observed in the MW.  This suggests that early type Oe stars are more common in low-metallicity environments.   

The frequency of Oe stars in each spectral type in the SMC is shown in Figure 4.  In the MW, the Oe/O ratio is estimated to be < 0.1-0.15 (Negueruela et al. 2004; Martayan et al. 2010).   
\articlefigure[scale=.45]{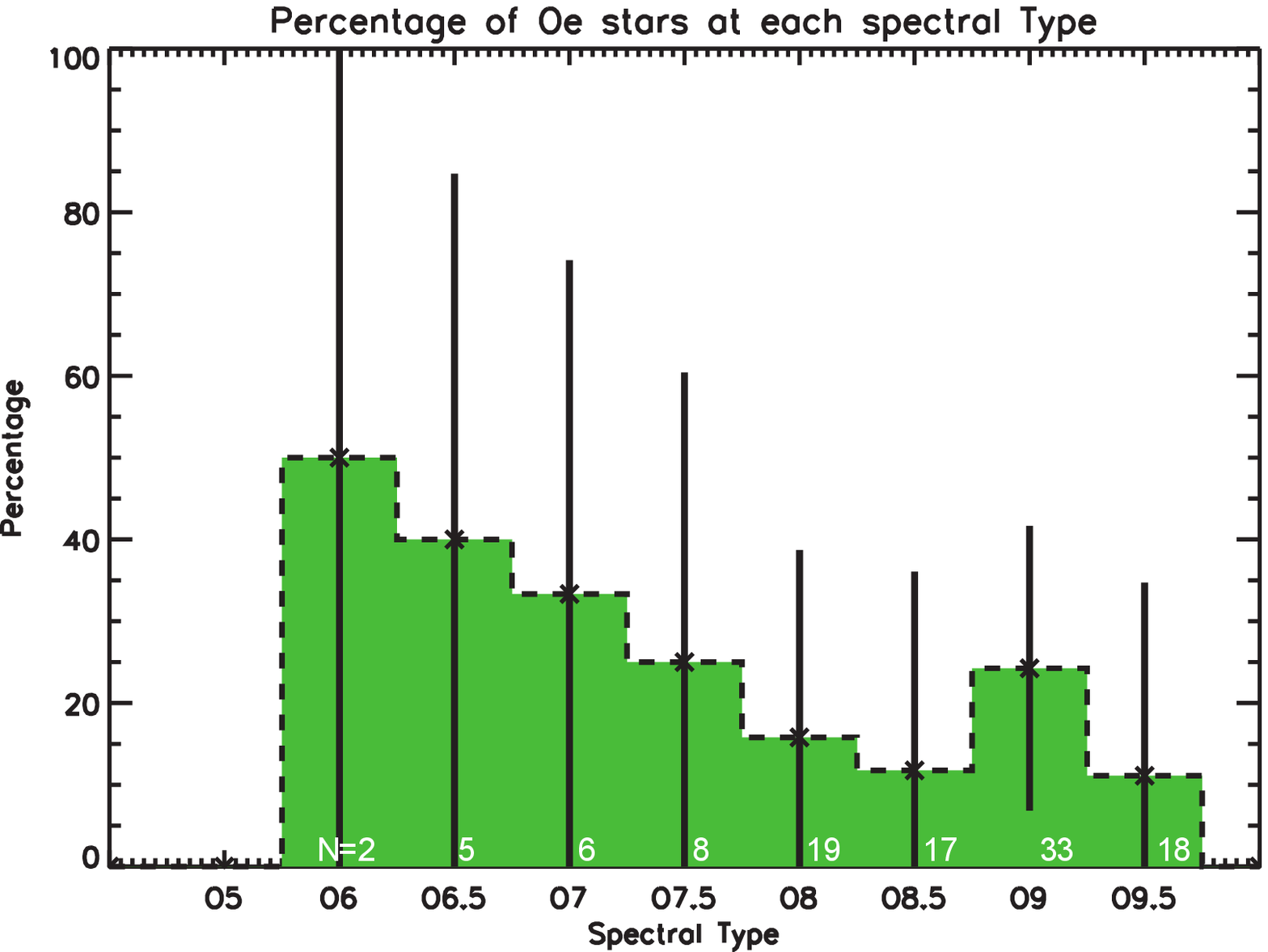}{ex_Figure 4}{The distribution of the percentage of O stars that are Oe stars at each spectral type.  For stars with uncertain classifications, late-type limits on the spectral types were used to compile this distribution.  The error bars represent the Poisson error in the total number, $N$, of O stars of each spectral type.  $N$ indicates the number of normal SMC field O stars plus SMC field Oe stars observed at each spectral type.} 
Based on Figure 4, the Oe phenomenon appears to be more frequent towards earlier spectral types in the SMC.  The frequency of Oe stars in each spectral type decreases from approximately 50\% to 20\% towards O8 -- O9 stars.  However, this trend may not be statistically significant since the error bars on the frequencies of the earliest type Oe stars are much larger than for the late-type stars.  The error bars come from the Poisson error in the total number of O stars of each spectral type.  We also estimate the frequency of all Oe stars relative to total O stars in the RIOTS4 survey, finding that 20 -- 28\% of field O stars are Oe stars.  

To compare the frequency of SMC field and cluster Oe stars, we evalute the frequency of O8e -- O9e stars, which have been observed in both the SMC field and clusters.  We find that 18 -- 21\% of our SMC field O8 -- O9 stars are Oe stars, while Martayan et al. (2010) find that in SMC clusters, 20 -- 23\% of all O8 -- O9 stars are Oe stars.  A comparison between these frequencies shows good agreement, confirming the contrast with the Milky Way, where the O8e -- O9e star frequency is about 14\% in clusters (Martayan et al. 2010), while hinting at the possibility that the frequency of Oe stars may be somewhat higher in clusters than in the field.   

\section{Future Work}

We have over 100 Be stars in the RIOTS4 survey that we will use to determine the spectral type distribution of SMC field Be stars.  We will use these stars to estimate how the effects of metallicity, and field versus cluster environment, differ between Oe and Be stars.  We also have multiple observations of 24 Oe/Be stars, which will allow us to study violet-to-red (V/R) variability in the line profiles, as well as variations in the strength of the Balmer  emission lines.  These may be linked to the density structure of the disk (e.g., Okazaki 1991) and the timescale of disk growth (e.g., Rivinius et al. 2013), respectively.  The RIOTS4 survey also includes many Oe/Be stars with Fe~\textsc{ii} emission lines, which are related to the decretion disk's structure and kinematics (Arias et al. 2006).  We have also initiated a spectroscopic monitoring survey of RIOTS4 stars using the Michigan Magellan Fiber System (M2FS), a high resolution, multi-object fiber system located on the Magellan Clay telescope (Mateo et al. 2012).  This high-resolution survey will allow us to measure more accurate rotation rates for our stars while also providing us with multiple observations to study V/R and Balmer line variability.  A major goal of our M2FS survey is to evaluate the binarity of the field OB stars, which may help to illuminate the role binarity plays in the Be phenomenon.   

\section{Conclusions}

We have identified the earliest classifiable Oe star to date, an O5.5-6 III((f))e star.  We found 25 -- 31 Oe stars in the RIOTS4 survey, which includes 5 -- 7 Oe stars of spectral type earlier than O7.5e, the earliest MW Oe star (Negueruela et al. 2004).  We find that 20 -- 28\% of SMC field O stars are Oe stars, and 20 -- 23\% of our Oe stars are of earlier spectral type than any observed MW Oe star to date.  Thus, we find that the frequency of Oe stars is higher in the SMC field than in the MW, where the frequency of cluster Oe stars is estimated to $\leq$ 10 -- 15\%  (Negueruela et al. 2004; Martayan et al. 2010).  We find that SMC field Oe stars are similar in frequency to that of SMC cluster stars evaluated by Martayan et al. (2010), and that this frequency is 2 -- 3 times greater in the low-metallicity SMC than in MW clusters. These results support the decretion disk model for Oe/Be stars, which predicts that the weaker stellar winds of O stars at low metallicity allow the retention of angular momentum and formation decretion disks, promoting both higher frequencies of Oe stars and earlier Oe spectral types.

\acknowledgments We thank the organizers of the Bright Emissaries Conference for the opportunity to present our research and for a wonderful venue and stimulating science program.  This work was supported by funding from NSF grant AST-0907758 to MSO.  Travel support was provided by the University of Michigan Rackham Graduate School.

\section{Comments from the Bright Emissaries Conference}

\underline{D. Baade}: Is anything known about the multiplicity of the Oe stars you found? \\
\\
\underline{Response}: While we have not yet studied the multiplicity of the Oe stars, our M2FS monitoring survey will allow us to look at the binary of the Oe stars.\\
\\
\underline{G. Wade}: There may be weak contamination of your Oe/Be samples by magnetic stars with glowing magnetospheres.\\
\\
\underline{S. Owocki}: I was intrigued that SMC Be stars may rotate at rates > 0.95 $\Omega_{\rm crit}$, in contrast to the MW rotation rate, 0.75-0.8 $\Omega_{\rm crit}$.  If true, I wonder if the reduced opacity makes it more difficult to excite pulsations in NRP stars, so the star only sheds excess angular momentum and ejects mass when it approaches closer to $\Omega_{\rm crit}$, while the higher metallicity stars can excite pulsations that make ejection possible around rotation rates of 0.75-0.8 $\Omega_{\rm crit}$.

\end{document}